Self-organization of gene regulatory network motifs enriched with short transcript's half-life transcription factors

Edwin Wang* & Enrico Purisima

Short title: Network motifs enriched with short transcript's half-life transcription factors

Biotechnology Research Institute, National Research Council Canada, Montreal, Quebec, Canada H4P 2R2.

Correspondence should be addressed to E.W. (Edwin.Wang@cnrc-nrc.gc.ca). Tel: 514-496-0914. Fax: 514-496-5143.

Teaser:

A common design principle of network motifs is that transcription factors with short transcript's half-lives are significantly enriched in motifs and hubs. This enrichment becomes one of the driving forces for the emergence of the network scale-free topology and allows the network to quick adapt to environmental changes.

Abstract

Network motifs, the recurring regulatory structural patterns in networks, are able to self-organize to produce networks due to the large ratio of genes to transcription factors (TFs) in genomes. The enrichment in motifs of the TFs with short transcript's half-lives can be seen as a motifs' common design principle and one of the driving forces for the emergence of the network scale-free topology, and allows the network to quickly adapt to environmental changes. Motifs are classified into subtypes that are preferentially used in different cellular conditions.

Gene regulatory networks are viewed as directed graphs, in which nodes represent transcription factors (TFs) and operons while the regulatory relationships are represented by edges. In these networks, three major kinds of motifs are observed: single input (SIM), bi-fan and feedforward loop (FFL)[1-3] (Figure 1). Network motifs can be seen as functional and structural units and the emergence of these motifs leads to the self-organization of the network[2,4]. By self-organization we mean that without the addition of extra connections, the links already present in the motifs define an extensive network that includes the majority of nodes in the entire network. We will illustrate this with the *E. coli* gene regulatory network using the known gene regulation data from a literature-mined database, RegulonDB[5] and other sources[6,7]. An explanation of more network terms used in this paper is included in Supplementary Notes.

In a network, a subset of links forms the network backbone which maintains the interconnections (directly and indirectly) between most TFs and thus maintains the integrity of the network. Without these backbone links, the graph would be fragmented into a collection of islands of smaller networks. How important are the motifs for maintaining the integrity of the network? If we remove all the motif links from the *E. coli* gene regulatory network, the network falls apart into disconnected islands (Supplementary Figure 1). Conversely, if we remove all links that are not part of motifs, we are left with a core network that preserves the backbone links (Supplementary Figure 1).

It is known that bi-fans are essential to maintain the network backbone links[8]. We examined further whether other motifs are also essential for network integrity. Removal of all FFL links did not destroy network integrity, while removal of all SIM links resulted in network fragmentation as did removal of all bi-fan links (Supplementary Notes). This suggests that bi-fans or SIMs are able to self-organize to form networks. This triggers the question as to how the motifs self-organize to form a network.

**The large ratio of genes to TFs in genomes results in self-organization of motifs**

When two motifs contain the same TF or gene, they self-assemble, i.e., they automatically form a networked pair of motifs (Supplementary Figures 2 and 3). The large gene regulatory network arises from the self-organization of motifs that share common TFs or genes. We tested this concept by simulating 496 bi-fans by randomly sampling the *E. coli* genome pool containing 116 TFs and 321 operons (Supplementary Notes). This is the same number of bi-fans found in the original network. Without adding any extra links, the motifs self-organized into a network. The self-organization occurred because of the limited number of distinct TFs relative to the number of bi-fans constructed in the simulation. We find that if the number of randomly generated bi-fans is larger than or equal to the number of TFs used to generate the bi-fans, these bi-fans are able to self-organize to form a network (Supplementary Notes). This is the situation in cells, where one TF often regulates many target genes, while one gene is regulated by many TFs. All genomes encode a limited number of TFs but a large number of regulated genes. Therefore, it is common for a genome to have more bi-fans than TFs, consequently leading to the self-organization of motifs into a large network.

Aside from being self-organized, the *E. coli* gene regulatory network is also scale-free[1,9]. Visually, scale-free networks are characterized by the presence of hubs in the network, i.e., nodes that are directly connected to a large number of other nodes. On the other hand, the randomly generated bi-fans, although they self-organize, do not form a scale-free topology due to the even distribution of TFs among the bi-fans. In the real gene regulatory network, the TFs in the bi-fans are very unevenly distributed. Randomization tests showed that the TF combinations in bi-fans are significantly different from chance expectation (<u>Supplementary Notes</u>). The same unevenness in TF pairs is also observed for the FFLs.

**Preferential usage of short transcript half-life transcription factors in hubs and network motifs**

The uneven TF combinations in FFLs and bi-fans are intriguing and it is tempting to speculate as to its origin. In an attempt to gain some insight into this, we first collected *E. coli* transcripts' half-lives from the Bernstein and coworkers[10] and mapped them onto the TFs in the network. Among the 116 TFs in the network, 107 of them have THLs mapped. They range from 14.4 to 1.9 minutes with a median of 5.5 minutes. Taking 5.2 minutes or less as a short THL, we calculated the percentage of short THL TFs in FFLs and bi-fans. Surprisingly we found that 60.0% of the FFL and bi-fan TFs have short THL. In contrast, 43.0% of the TFs in the network have short THLs (Table 1). Furthermore, 94.5% of the FFL TF pairs and 92.0% of the bi-fan TF pairs contain at least one short THL TF (<u>Supplementary Tables 2 and 3</u>). We also confirmed that this phenomenon is not by chance (<u>Supplementary Notes</u>). Extending this analysis to SIMs, we find about 70.0% of SIM TFs have short THLs (Table 1). These data show that there is a preferential selection of short THL TFs in SIMs, FFLs and bi-fans. This preferential selection will lead to an uneven numbers of target genes regulated by each TF and therefore become one of the driving forces to generate a scale-free network topology, i.e., it leads to the formation of hubs in the network. In fact, about 70.0% of hub nodes are TFs with short THLs ($p < 0.05$, Table 1). It has been previously reported that FFLs and bi-fans are naturally selected[11]. Here we find that short THL TFs become a selection trait in hubs and all the motifs (SIM, FFL and bi-fan).

Short THL TFs can alter their transcript concentrations quickly, which will facilitate the motifs' adaptation to rapid condition changes[12]. Short THL TFs also mitigate gene expression fluctuations, or internal noise[13,14], which can garble cell signals and corrupt circadian clocks[15]. Taken together, the network generated by self-organization of these motifs has evolved to be more robust and adaptable to the cellular condition changes. The preferential usage of short THF TFs in hubs, and network motifs allows gene expression to turn on and off quickly, which represents a common design principle of these motifs and the network. The frequent occurrence of FFL and bi-fan TF pairs containing one short THL TF can be seen as a criterion for self-organizing FFLs and bi-fans.

**Motif and hub subtypes and their usages in different cellular conditions**

To explore the relationship between the motifs and TF's THLs, we classified the motifs and hubs into 2 types based on the THLs of their TFs (Table 2). The classification and the characteristics of these subtypes are discussed in Supplementary Notes. To get a dynamic view of the motif subtype usages in different cellular conditions, we reconstructed 3 sub-networks using 22 microarray experimental data of the *E. coli* grown in these conditions: logarithmic growth phase, diauxic shift and the stationary phase (Supplementary Notes).

Table 2 summarizes the dynamic representation of the networks and the motif subtype usages. A more detailed discussion of the results is included in Supplementary Notes. Briefly, the frequencies of Type I FFLs in the three sub-networks are similar, suggesting that FFLs may be used as buffers to maintain some biological processes. Type I hubs, bi-fans and SIMs are favored by an active growth condition in which many biological processes are coordinated and quickly respond to the inducing conditions. On the other hand, the Type II bi-fans are favored by the cellular conditions such as the stationary and diauxic shift, which significantly reduce the level of the biosyntheses of DNA and protein and inhibit aerobic metabolism as reported previously.

## Concluding remarks

In conclusion, the large ratio of genes to TFs in genomes leads to a sharing of TFs or genes by motifs and is sufficient to result in their self-organization. The enrichment of short THL TFs in motifs and hubs allows the network to quickly adapt to environmental changes, which represents a common design principle of the motifs. Furthermore, it becomes one of the driving forces for the emergence of the network scale-free topology. Most FFLs and bi-fans contain at least one short THL TF, which can be seen as another criterion for self-assembly of these motifs. We have classified the motifs according to their short THL TF content. We show that the percentage of the different motif subtypes is dependent on the cellular conditions.

## Acknowledgments

We thank J. Collado-Vides and his colleagues for providing the data of the RegulonDB database. We thank M.M. Babu at LMB-MRC for many useful suggestions regarding the manuscript that led to the interpretation of hubs having short transcript half-lives. We also thank M. Whiteway, P. Lau, H. Hogues, Z. Yu, C. Wu and A. Nantel for discussions. This work is supported by Genome and Health Initiative (GHI), Canada. NRC publication No. 46238.

Table 1. The enrichment of the short transcript's half-life transcription factors in hubs and network motifs.

|  | TF | Natural rate (%) | Random rate (%) | P value |
|---|---|---|---|---|
| Hub | 11, 10 | 70.0 | 43.1 | < 0.05 |
| SIM | 13, 10 | 70.0 | 43.2 | < 0.05 |
| FFL | 29, 21 | 61.9 | 42.9 | < 0.04 |
| Bi-fan | 38, 28 | 60.1 | 43.1 | < 0.03 |

In the second column, the first number represents the total number of transcription factors (TFs) in each component (hub, SIM, FFL and bi-fan). The second number represents the number of TFs out of the first that have a mapped transcript's half-life (THL). In the third column, the natural rate represents the observed fraction of the TFs having short transcript half-lives (THLs) in the THL-mapped TFs. When a TF's THL is equal to or shorter than 5.2 min, we say this TF is a short THL TF. In the fourth column, the random rate represents the fraction of the short THL TFs in the randomly sampled TFs. P value represents the probability. Details in statistical analysis are included in Supplementary Notes.

Table 2. Dynamic representation of gene regulatory networks and the frequencies of motif types in different cellular conditions.

| | | Static | Aerobic | Diauxie | Stationary |
|---|---|---|---|---|---|
| Size | No. of transcription factors | 116 | 61 | 84 | 35 |
| | No. of target operons | 321 | 201 | 229 | 96 |
| | No. of regulatory interactions | 567 | 300 | 372 | 147 |
| Type I hubs | Hub | 11 (70.0%) | 6 (100.0%) | 8 (71.4%) | 3 (66.7%) |
| Type I motif | SIM | 13 (70.0%) | 11 (91.7%) | 9 (69.2%) | 5 (71.4%) |
| | FFL | 14 (77.8%) | 4 (80.0%) | 7 (77.8%) | 3 (100.0%) |
| | Bi-fan | 20 (40.0%) | 11 (68.8%) | 8 (38.1%) | 3 (37.5%) |

Changes in the gene regulatory networks are tabulated for the static, logarithmic growth, diauxic shift (diauxie) and stationary phases. The last four rows show the number and fraction of hubs and motifs active in the various cellular conditions. Type I FFLs in which the first transcription factor (TF) transcript's half-life (THL) is shorter than the second one's, and Type II FFLs in which the first TF's THL is longer than the second one's. For bi-fans, hubs and SIMs, if a TF's THL is 5.2 min or less we define it as a short THL TF. A bi-fan is defined as Type I if both TFs have short THLs and as Type II if one of the TF pair has a long THL. Hubs or SIMs are defined as Type I if they have short THL TFs and Type II if they long THL TFs.

Figure legends:

Figure 1. Network motifs in the *E. coli* gene regulatory network.

**A,** Single input module (SIM): a transcription factor (TF) regulates a group of genes (G1, G2, G3 and G4). **B,** Feedforward loop (FFL): a transcription factor (TF1) regulates the second transcription factor (TF2), both TF1 and TF2 regulate a target gene (G1). **C,** Bi-fan: both transcription factors TF1 and TF2 regulate both target genes (G1 and G2). In the *E. coli* gene regulatory network, we identified 58 FFLs, 496 bi-fans and 13 SIMs.

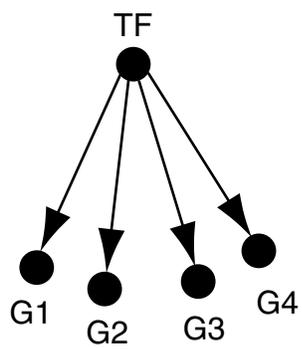 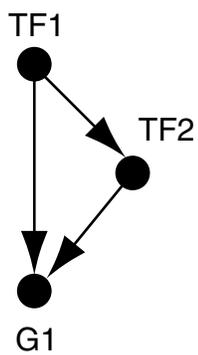 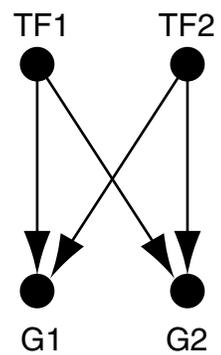

A
B
C

Figure 1

**Supplementary Notes**

**Self-organization of gene regulatory network motifs enriched with short transcript's half-life transcription factors**

Edwin Wang* and Enrico Purisima

Biotechnology Research Institute, National Research Council Canada, Montreal, Canada H4P 2R2.

*Correspondence should be addressed to:
Email: Edwin.Wang@cnrc-nrc.gc.ca
Tel: 514-496-0914
Fax: 514-496-5143

# Contents

**A. Datasets**

**A1. Datasets for the *E. coli* gene regulatory network.** For constructing the *E. coli* gene regulatory network we combined the data from RegulonDB[1] and other sources[2,3]. All the sigma factors were removed from the datasets, because most of them are associated with the core RNA polymerase complex to initiate gene transcription, and do not directly regulate gene expression. The *E. coli* gene regulatory network is represented as a directed graph, in which nodes represent transcription factors (TFs) and operons while the regulatory relationships are represented by edges. The network contains 437 nodes (116 transcription factors, 321 operons and ~900 genes) and 567 edges. In this paper, we refer to this network as a static network. Networks were visualized using Pajek (http://vlado.fmf.uni-lj.si/pub/networks/pajek/).

**A2. Microarray data for constructing cellular sub-networks.** We compiled 22 *E. coli* microarray time series data for these cellular conditions: logarithmic growth in aerobic condition (www.ou.edu/microarray/oumcf/glucosetime.htm), diauxie and stationary phase (www.ou.edu/microarray/oumcf/diauxietime.htm).

**A3. Transcript half-lives of the transcription factors in *E. coli*.** The *E. coli* gene transcript half-lives (THLs) data were obtained from Bernstein and coworkers[4]. We took the genes' transcript half-lives for bacteria cultured on M9 medium supplemented with glucose as a carbon source[4].

**B. Terms in gene regulatory networks**

We list some network terms in the context of gene regulatory networks. More detailed or expanded explanations of these terms can be found in these papers[5,6].

In gene regulatory networks:

***node*** represents transcription factors (TFs) or genes/operons.

***In-degree*** of a gene/operon represents the number of TFs which regulate the gene/operon; a higher in-degree of a gene/operon indicates that this gene is regulated by more TFs.

***Out-degree*** of a TF represents the number of genes/operons which are regulated by the TF; a higher out-degree of a TF means that this TF regulates more genes/operons in the network.

***Out-degree distribution*** of a gene regulatory network gives the probability that a selected TF can regulate the number of genes/operons.

***Hub:*** in a gene regulatory network, there are a few high out-degree TFs that regulate many genes/operons, we called these TFs hubs. Normally hubs are global TFs.

***Scale-free network:*** if the distribution of the out-degree or the in-degree of nodes in a network follows a power-law, we called the network scale-free network. In terms of biology, scale-free gene regulatory networks are characterized by the presence of a few hubs in the network, i.e., a few global TFs regulate many target genes. Visually, in networks, these global TFs (called hubs) are directly connected to a large number of other nodes (regulated genes/operons).

***Clustering coefficient*** of a network measures the inter-regulations between TFs; a higher clustering coefficient tells more inter-regulations between TFs.

***Network motifs*** are the statistically significant recurring regulatory structural patterns that are present in a real network in contrast to randomized networks. Three major motifs are found in gene regulatory networks: single input Module (SIM), bi-fan and feedforward loop (FFL). The SIM has one TF that regulates many target genes. The bi-fan has two TFs that together regulate two target genes. The FFL has one TF that regulates a second TF and both directly regulate a target gene.

***Network backbone*** in gene regulatory networks is a subset of links that maintain the interconnections (directly and indirectly) between most TFs and then maintains the integrity of the network.

***Network formation by self-organization of network motifs***: we mean that without the addition of extra connections, the links already present in the motifs define an extensive network that includes the majority of nodes in the entire network. This can be happen because when two motifs share TFs or target genes, these two motifs can be self-assembled together.

**C. Reconstruction of the sub-networks of different cellular conditions.** To construct the sub-networks of different cellular conditions, we modified the trace-back algorithm developed for constructing yeast cellular condition-specific sub-networks[6]. We first constructed the aerobic sub-network as follows: we identified the modulated TFs and genes (up- or down-regulated) of the each time point using the Iterative Group Analysis method[7] for the first 6 time points of the logarithmic growth phase data which covered the length of the logarithmic growth phase of the cells. We combined all the identified TFs and genes and then mapped each gene to the operon which contained the gene. To get a sub-pool of the TF and operon regulation relations, we queried the TFs and operons to find out all the TF and opreron regulation relations documented in the static network, and then removed the $TF_1$-$TF_2$ ($TF_1$ regulates $TF_2$) regulation relations if the $TF_2$ did not regulate any genes. Using this sub-pool, we produced the sub-network. The same procedure was applied to construct other sub-networks. We used the first 12 and last 4 time points of the diauxie microarray data to construct the diauxie and stationary phase sub-networks, respectively.

**D. Network motif and hub identification.** The algorithms used for detecting FFL and SIM motifs are the same as those described previously[8]. To determine the minimal number ($N_{sim}$) of the target genes for a motif to be classified as a SIM, we divided the total number of the regulation interactions by the total number of TFs in the network. A motif is a SIM if the TF regulates more than $N_{sim}$ operons. In the static network, $N_{sim}$ is 9. $N_{sim}$ should vary in different networks. To detect the bi-fan motifs, we constructed a matrix $\boldsymbol{M}$, in which all the TFs and operons were in columns and the TFs in rows. $M_{ij} = 1$ if row $i$ regulates column $j$. Otherwise, $M_{ij} = 0$. We also set $M_{ij} = 0$ if $i$ and $j$ represent the same TF, i.e., self-regulation. For each pair of rows $i_1$ and $i_2$ we create a list consisting of all $j$ for which $M_{i1j} = M_{i2,j} = 1$, i.e., a list of all TFs or operons simultaneously regulated by the TFs $i_1$ and $i_2$. If the size of this list is greater than 1, then one or more bi-fans can be constructed by taking the TFs $i_1$ and $i_2$ along with all pairs of entries in the list. In this manner, we exhaustively enumerate all bi-fan motifs in the network.

To determine the hubs, we ranked all the TFs based on the number of the target operons regulated by each TF in a descending order. We took the top 10% of the ranked TFs as hubs.

**E. Topological characteristics of the removal of the motif links from the static network.** We removed the links of the FFLs, SIMs, bi-fans, or the links of all the motifs, respectively (we called the resulting networks as target-removed networks: target-FFL network, target-SIM network, target-bi-fan network and target-all network). As a control, we randomly removed links equal in number to the links of all the motifs (we called the resulting network as a randomly-removed network). For each network, we examined the number of nodes in the largest connected component and the out-degree distributions of the TFs. To obtain good statistics on the randomly removed network, 1000 simulations of the randomly removed network were carried out and the node numbers of the most connected components were counted. In each simulation, a similar result was observed. For each of the target- and randomly- removed network, we also examined other general network topological measures using the statistical analysis of network dynamics method (SANDY)[6]. The results are shown in Supplementary Table 1. The graphs in Supplementary Figure 1 provided a visual summary about the target removal of the links of bi-fans, SIMs or all the links of the motifs.

In the target-removed networks, the ratios of the node number of the largest connected component to the resulting networks (fc in Supplementary Table 1) change significantly. The ratio is the highest in the static and target-FFL networks and the lowest in the target-all network. In contrast, the ratio in the randomly-removed network is higher than that of the target-all network. In the target-bi-fan and the target-SIM networks, the ratios are also low. The ratio (fc) is a measure of the connectivity of the network; a higher ratio in a network means that the network is largely connected. These results suggest that bi-fans and SIMs are essential to maintain the integrity of the network, however, FFLs are not. This is also consistent to this fact: the average of the out-degree of the TFs ($k_{out}$ in Supplementary Table 1) in the target-FFL network is similar to the static network and higher than other target-removed networks. This suggests that targeted removal of FFL links did not break many links between TFs and operons. One reason is that the number of the FFLs is relatively small; the other reason is that most FFL TFs also take part in bi-fans or SIMs. Therefore, targeted removal of the FFL links cannot affect TFs' out-degrees and the network integrity. The clustering coefficients (cc in Supplementary Table 1) in the target-FFL and the target-all networks are zero but not in other networks, suggesting that FFLs provide TF inter-regulations in the network. Finally, the out-degree distributions of TFs ($\gamma$ in Supplementary Table 1) of all the networks follow a power-law, suggesting these networks are still scale-free networks; however, the out-degree distributions of TFs vary in the target-removed networks. In the target-SIM and the target-all networks, the $\gamma$ values are higher than 3. When the $\gamma$ value is greater than 3, most properties of the scale-free network are lost[5]. Therefore, targeted removal of the SIM links or all the motif links leads to the network collapse and the loss of the scale-free network properties.

**F. Motifs are self-organized to form networks by sharing transcription factors or operons.** If two or more motifs contain the same TF or the same target gene, these motifs are self-organized or self-assembled. To illustrate the network formation by self-organizing motifs, we constructed 3 motif-based networks, FFL-network, SIM-network and bi-fan-network using FFLs, SIMs and bi-fans, respectively. Supplementary Figure 3 shows the graphs of these three networks and the sharing events of operons and TFs.

**G. Self-organization of the randomly generated bi-fans.** In the original network, there are 116 TFs and 321 operons. To construct randomized bi-fans, we randomly sampled the TFs and operons from a genome pool containing 116 TFs and 321 operons, while assuming TFs have in- and out- degrees and operons have in-degrees only. We randomly sampled 2 TFs (tf1 and tf2) and 2 operons (g1 and g2) at the same time, and then wired them in such a way that tf1 becomes connected to g1 (tf1→g1), tf1 to g2 (tf1→g2), (tf2→g1) and tf2 to g2 (tf2→g2). In each simulation, we constructed 496 randomized bi-fans, the same number as in the original network.

Motifs are self-organized by sharing common TFs or target genes. When we put these randomized bi-fans together without adding any other links, they formed a random network with evenly distributed out-degrees for each TF. However, a scale-free topology was not found in these networks formed by randomized bi-fans.

**H. Mapping the transcript half-lives to the network TFs.** We mapped the transcript half-lives (THLs) to the TFs extracted from the *E. coli* static network. The absolute values of the THLs of the genes in bacteria may change under different conditions, but their relative values do not change too much[9]. Among 116 TFs in the network, 107 of them have been mapped to their THLs. The longest value of the TF's THL is 14.4 min, the shortest is 1.9 min and the median is 5.5 min. The trimmed mean ($\gamma$=0.20) of this dataset is 5.4 min, and the Bootstrap-t 95% confidence interval of 50000 times bootstrap is from 5.2 to 6.1 min. If the THL of a TF is less than 5.2 min or equal to/greater than 5.2 min, we consider this TF as having a short or long THL, respectively. We mapped the THLs to the TF pairs in motifs. For bi-fans and FFLs which have two TFs in each circuit, we considered the THLs of the motifs' TFs as mapped, if both TFs in a circuit have been mapped to their THLs. In total, 18 and 50 TF combination pairs of the FFLs and bi-fans were mapped to their THLs, respectively (Supplementary Tables 2 and 3). About 94.5% of the FFL TF pairs and 92.0% of the bi-fan TF pairs contained at least one short THL TF.

**I. Subtypes of hubs and network motifs.** To explore the relationship between the motifs and TF's THLs, we classified the motifs based on the THLs of their TFs. The FFL circuits can be divided into 2 types: Type I FFLs in which the first TF's THL is shorter than the second one's, and Type II FFLs in which the first TF's THL is longer than the second one's. We find that in whole network 77.8% of the TF pairs belong to Type I while 22.2% of the TF pairs belong to Type II (Table 2 in the main text).

The fast decay of a short THL TF offers the ability to alter transcript concentrations rapidly, allowing a quick response to rapid condition changes in an input signal. A longer THL TF can maintain its concentration relatively longer and filters out short-term fluctuations or internal noise and provides a stable biological response. Besides, most FFLs can function as sign-sensitive delay circuits[10,11].

We define a bi-fan as Type I if both TFs have short THLs and as Type II if one of the TF pair has a long THL. We find that 40% of bi-fans are Type I bi-fans while 60.0% are Type II. Bi-fans normally integrate 2 input signals to regulate the target genes[12]. Most short THL TFs are global regulators or regulators for the machinery of DNA and amino acid biosynthesis, while most of long THL TFs are the regulators involving catabolic repression. Type I bi-fans are favored in rapidly integrating external signals and DNA/amino acid biosynthesis, whereas Type II bi-fans can maintain certain negative feedback such as inhibiting aerobic metabolism.

Most hubs and SIMs have short THL TFs. We will refer to these as Type I hubs or Type I SIMs. The advantage of Type I hubs or SIMs is the ability for a rapid large-scale global and coordinated response to a single input signal[13]. This is due to the central nature of hubs and SIMs which directly regulate multiple genes. Therefore, most hubs or SIMs could integrate a few different biological processes together and rapidly respond to a single input signal in a large-scale and a coordinated manner.

**J. Motif and hub subtype usages in different cellular conditions.** To get a dynamic view of the motif subtype usages in different cellular conditions, we reconstructed 3 sub-networks using 22 microarray experimental data of the *E. coli* grown in these conditions: logarithmic growth phase, diauxic shift and the stationary phase (Supplementary A2 and Supplementary C). Cells are reprogrammed to inhibit aerobic metabolism and shut down most of their protein and DNA syntheses in diauxic shift[14]. DNA, protein and energy metabolisms are actively induced in the logarithmic growth phase. Most biological processes are shut down in stationary phase.

Table 2 in the main text summarizes the dynamic representation of the networks and the motif subtype usages. The sizes of the sub-networks vary significantly across the different phases. The smallest sub-network belongs to the stationary phase, indicating that many biological processes are shut down in this phase. Hubs and SIMs are the most important transcription factors in the network because they influence many target genes. In fact, SIM TFs are a superset of the hubs. Only one hub, crp is present in three sub-networks. We refer to it as a permanent hub. Crp is known as a global regulator and senses cAMP concentration, which rises during glucose starvation and decreases during growth on glucose. The percentage of Type I SIMs and hubs is approximately 90% for the aerobic sub-network and with a somewhat lower value of 70% for the diauxic shift and stationary phases. Similarly, for bi-fans, we see a significantly higher frequency (68.8%) of Type I circuits in the aerobic sub-network than in the stationary and diauxic shift sub-networks (~ 38.0%). These data suggest that Type I hubs, bi-fans and SIMs are favored by an active growth condition in which many biological processes are coordinated and quickly respond to the inducing conditions. On the other hand, the Type II bi-fans are favored by the cellular conditions such as the stationary and diauxic shift, which significantly reduce the level of the biosyntheses of DNA and protein and inhibit aerobic metabolism as reported previously. For example, one of the hubs/SIMs in stationary and diauxic shift sub-networks is a long THL TF, fnr (THL, 6.8 min), which repress aerobic metabolism.

The frequencies of Type I FFLs in the three sub-networks are similar. In these three conditions, FFLs may be used as buffers to maintain some biological processes. For example, a Type I FFL containing the crp (HTL, 3.2 min)-caiF (HTL, 10.1 min) pair, is present across the three sub-networks. The long THL TF, caiF is known to regulate and maintain multipurpose metabolic conversions.

## K. Randomization tests

**K1. Randomization tests for the occurrence of the transcription factor pairs in motifs.** To test the statistical significance of the uneven occurrence of TF pairs in bi-fans and FFLs, we first constructed randomized motifs. To construct randomized motifs, we used the same procedures described previously. We kept the number of randomized FFLs and bi-fans equal to those in the static network; this means that we constructed 58 FFL and 496 bi-fan randomized motifs in each test.

We first calculated the natural distribution rate of a TF combination pair in both motif types. We defined $P_{tf1\text{-}tf2|motif}$ as the natural distribution rate of a TF combination pair in a particular motif type, which is calculated as the percentage of motifs of that type containing the TF pair. We found that the distribution rates of the TF pairs in both motif types were not evenly distributed. To test the statistical evidence of the distribution rates of the TF pairs in FFLs or bi-fans, we defined $P_{tf1\text{-}tf2|motif\_random}$ as the random distribution rate of a TF pair in a randomly constructed particular motif type, which is calculated as the percentage of randomly constructed motifs of that type containing the TF combination pair. We tested the null hypothesis $P_{tf1\text{-}tf2|motif\_random} \geq P_{tf1\text{-}tf2|motif}$ by performing 10000 resampling tests. When we performed the tests for FFLs, the order of two TF combinations was considered, for example, we distinguished these two situations in FFL circuits: (1) in one circuit, Gene A as the first TF that regulates Gene B, the second TF; (2) in another circuit, Gene B as the first TF that regulates Gene A, the second TF. When we performed the tests for bi-fans, the order of two TF combinations was ignored. All p values of the multiple tests were adjusted by the false discovery rate (FDR) method. We find that TF pairs are unevenly distributed in bi-fans and FFLs ($p < 10^{-3}$),

**K2. Randomization tests for the fractions of short transcript half-life transcription factors in the TF pairs of the bi-fans and FFLs.** To test whether the fractions of the short transcript half-life (sTHL) TFs in the TF pairs of the bi-fans and FFLs are by random, we performed randomization tests. We used 2 sample spaces: in the first set of the tests, we used the total THL mapped TFs (107 TFs) in the static network as a sample space. To test the statistical significance of the sTHL TFs in bi-fans, we defined $P_{tf1\text{-}tf2}$ as a fraction of TF combination pairs containing at least one sTHL TF and $P_{tf1\text{-}tf2|decay\_Rand}$ as a fraction of the randomized TF combination pairs containing at least one sTHL TF, while ignoring the order of the 2 TFs. To construct the randomized TF combination pairs, we randomly sampled 2 TFs from the pool containing the 107 TFs, which were mapped with their THLs. We tested the null hypothesis $P_{tf1\text{-}tf2|decay\_Rand} \geq P_{tf1\text{-}tf2}$ by performing 10000 times of resampling tests. In each test, we constructed 50 TF combination pairs. We rejected the hypothesis if the $p < 0.05$. We find that the TF combination pairs having at least one sTHL TF are not by chance ($p < 10^{-3}$). The same procedure was applied to the tests for FFLs. We constructed 18 TF pairs in each test and a similar result was obtained. In the second set of the tests, we used the THL mapped TFs which were found in either bi-fans or FFLs as a sample pool. In this set of the tests, we tested whether the observed fractions of the sTHL TFs in the TF pairs are due to the distributions of the TFs in the

motifs. Particularly, we found that the sTHL TFs were enriched in bi-fans and FFLs, respectively (see the main text and Supplementary Table 5). We asked if this enrichment led to the observed fractions of the sTHL TFs in the TF pairs of the bi-fans or FFLs. To answer this question, we performed the second set of the tests. The test procedures were the same as the first set of the tests except using a different sample space, which was a TF pool consisting only of those found in the motifs. The results of these tests are summarized in Supplementary Table 4.

**K3. Randomization tests for the fractions of the short transcript half-life transcription factors in hubs and motifs.** We tested the statistical significance of the enrichment of the sTHL TFs in hubs and motifs. We used the 107 TFs, which were mapped with their THLs in the static network, as a resampling pool. We defined the observed percentage of the sTHL TFs in SIMs as $P_{sim.}$ In each resampling, we randomly took 13 TFs and calculated the fraction of sTHL TFs ($P_{rand}$) in these 13 TFs. When a TF's THL is equal to or shorter than 5.2 min, we say this TF is a sTHL TF. We tested the null hypothesis $P_{rand} \geq P_{sim}$ by performing 10000 times of resampling tests. We rejected the hypothesis if p < 0.05. The same procedure was applied to test the fractions of the sTHL TFs in FFLs, bi-fans and hubs. The results of these tests are summarized in Supplementary Table 5.

**K4. Randomization tests for the distributions of Type I hubs and motifs (SIM, FFL and bi-fan).** Randomization tests based on the above definitions have been carried out. The tests for Type I hubs and SIMs are described in Supplementary K3 and the results shown in Supplementary Table 5. For Type I bi-fans and FFLs, we performed similar tests as described in Supplementary K2. We also used the same 2 sample spaces in K2. For the randomized TF combination pairs for the FFL tests, we took the order of the TFs in consideration. The results are shown in Supplementary Table 4.

**Supplementary Table 1: Network motifs, bi-fans and SIMs but not FFLs are essential to maintain the integrity of the network**

| Network | $k_{in}$ | $k_{out}$ | cc | fc (%) | $\gamma$ |
|---|---|---|---|---|---|
| Static network | 1.65 | 4.89 | 0.21 | 80.59 | 2.38 |
| Target-FFL network | 1.45 | 4.14 | 0.00 | 71.93 | 2.21 |
| Target-bi-fan network | 1.18 | 3.02 | 0.15 | 32.40 | 2.62 |
| Target-SIM network | 1.16 | 2.27 | 0.10 | 9.10 | 3.07 |
| Target-all network | 1.05 | 1.75 | 0.00 | 4.30 | 3.38 |
| Randomly-removed network | 1.17 | 2.89 | 0.16 | 44.90 | 2.14 |

$k_{in}$ represents the average in-degree of operons and TFs. $k_{out}$ represents the average out-degree of TFs. cc represents the average clustering coefficient of the network. fc represents the percentage of nodes contained in the largest connected component (the largest subgraph in the network). $\gamma$ represents the out-degree distribution of the TFs in the network.

**Supplementary Table 2: Transcript half-lives of the transcription factor pairs in feed-forward loops (FFLs)**

| 1st transcription factor | Half-life (min) | 2nd transcription factor | Half-life (min) |
|---|---|---|---|
| crp | 3.2 | araC | 4.9 |
| crp | 3.2 | caiF | 10.1 |
| crp | 3.2** | fur | 2.4 |
| crp | 3.2 | galS | 4.8 |
| crp | 3.2 | glpR | 3.9 |
| crp | 3.2 | malI | 12.2 |
| crp | 3.2 | malT | 9.1 |
| crp | 3.2 | melR | 6.4 |
| crp | 3.2 | ompR | 14.4 |
| fnr | 6.8** | arcA | 2.6 |
| fnr | 6.8** | narL | 4.3 |
| gcvR | 3.3 | gcvA | 4.2 |
| hns | 3.6 | flhC | 5.1 |
| hns | 3.6 | flhD | 6.7 |
| metJ | 3.6 | metR | 5.7 |
| nac | 4.9 | putA | 8.1 |
| rob | 3.0** | arcA | 2.6 |
| ihf | 6.6* | ompR | 14.4 |

The first and the third columns represent the FFL's first and the second transcription factors, respectively. The second and the fourth columns list the transcript's half-lives for the transcription factors in the first and the second columns, respectively. One out of 18 transcription factor pairs (5.6%, marked by *) has both transcription factors with longer transcript half-lives. Four out of 18 transcription factor pairs (22.2%, marked by **) have the pattern that the transcript half-life of the first transcription factor is longer than the second one's.

**Supplementary Table 3: Transcript's half-lives of the transcription factor pairs in bi-fans**

| Transcription factor | Half-life (min) | Transcription factor | Half-life (min) |
|---|---|---|---|
| arcA | 2.6 | betI | 5.4 |
| arcA | 2.6 | crp | 3.2 |
| arcA | 2.6 | fis | 3.3 |
| arcA | 2.6 | fnr | 6.8 |
| arcA | 2.6 | ihf | 6.6 |
| arcA | 2.6 | marA | 5.0 |
| arcA | 2.6 | narL | 4.3 |
| arcA | 2.6 | rob | 3.0 |
| arcA | 2.6 | soxS | 5.3 |
| crp | 3.2 | araC | 4.9 |
| crp | 3.2 | caiF | 10.1 |
| crp | 3.2 | cytR | 4.5 |
| crp | 3.2 | deoR | 5.8 |
| crp | 3.2 | fis | 3.3 |
| crp | 3.2 | fnr | 6.8 |
| crp | 3.2 | fucR | 4.8 |
| crp | 3.2 | galR | 5.6 |
| crp | 3.2 | glpR | 3.9 |
| crp | 3.2 | gntR | 4.1 |
| crp | 3.2 | hns | 3.4 |
| crp | 3.2 | ihf | 6.6 |
| crp | 3.2 | malt | 9.1 |
| crp | 3.2 | mlc | 5.5 |
| crp | 3.2 | nagC | 2.4 |
| crp | 3.2 | narL | 4.3 |
| crp | 3.2 | ompR | 14.4 |
| crp | 3.2 | rhaS | 4.9 |
| crp | 3.2 | rhaR | 8.3 |
| csgD | 4.7 | ompR | 14.4 |
| cytR | 4.5 | deoR | 5.8 |
| dnaA | 1.9 | fis | 3.3 |
| fhlA | 3.0 | ihf | 6.6 |
| fis | 3.3 | fnr | 6.8 |
| fis | 3.3 | ihf | 6.6 |
| hns | 3.6 | fnr | 6.8 |
| hns | 3.6 | flhC | 5.1 |
| hns | 3.6 | flhD | 6.7 |
| metJ | 3.6 | metR | 5.7 |
| nac | 4.9 | putA | 8.1 |
| narL | 4.3 | ihf | 6.6 |
| narL | 4.3 | fnr | 6.8 |
| rob | 3.0 | fnr | 6.8 |
| rob | 3.0 | marA | 5.0 |
| rob | 3.0 | soxS | 5.3 |
| marA | 5.0 | soxS | 5.3 |
| tyrR | 3.3 | ihf | 6.6 |
| tyrR | 3.3 | trpR | 6.1 |
| ihf | 6.6* | envY | 7.9 |
| ihf | 6.6* | fnr | 6.8 |
| ihf | 6.6* | ompR | 14.4 |
| soxS | 5.3 | fnr | 6.8 |

The first and the third columns represent two transcription factors that are present in bi-fans. The second and the fourth columns list the transcript half-lives for the transcription factors in the first and the second columns, respectively. Three out of 50 transcription factor pairs (6.0%, marked by *) have both transcription factors with longer transcript half-lives.

**Supplementary Table 4: The relations of the transcription factors' transcript half-lives in bi-fans and feedforward loops (FFLs)**

|  | Relation | Natural rate (%) | Random rate (%) | P value |
|---|---|---|---|---|
| FFL, Pool 1 | TF1 or TF2 (short) | 94.5 | 67.6 | $< 10^{-3}$ |
|  | TF1 < TF2 | 77.8 | 48.7 | $< 10^{-3}$ |
| FFL, Pool 2 | TF1 or TF2 (short) | 94.5 | 82.3 | $< 0.04$ |
|  | TF1 < TF2 | 77.8 | 47.5 | $< 10^{-3}$ |
| Bi-fan, Pool 1 | Either TF (short) | 92.0 | 67.8 | $< 10^{-3}$ |
|  | Both TFs (short) | 40.0 | 19.9 | $< 10^{-3}$ |
| Bi-fan, Pool 2 | Either TF (short) | 92.0 | 85.1 | $< 0.05$ |
|  | Both TFs (short) | 40.0 | 37.8 | $< 0.67$ |

In the first column, FFLs, bi-fans and sample spaces are listed. Pool 1 and Pool 2 represent the 2 sample spaces: the total transcript half-life (THL) mapped TFs (107 TFs) in the static network and the THL mapped TFs found in each motif, respectively. In the second column, the relation represents the transcript half-life (THL) relationships between a transcription factor (TF) pair. "TF1 or TF2 (short)" in FFLs indicates that the transcript half-life (THL) of either TF1 or TF2 is equal to or shorter than 5.2 min. "TF1 < TF2" indicates that the first TF's THL is shorter than the second one's. In bi-fans, "Either TF (short)" indicates that the THL of either TFs is equal to or shorter than 5.2 min. "Both TFs (short)" indicates that the THLs of both TFs are equal to or shorter than 5.2 min. In the third column, the natural rate represents the observed fraction of the TF pair relation in TF pairs. In the fourth column, the random rate represents the fraction of the TF pair relation in the randomly sampled TF pairs. P value represents the probability.

**Supplementary Table 5: The enrichment of the short transcript half-life transcription factors in hubs and motifs (SIM, FFL and bi-fan)**

|        | TF     | Natural rate (%) | Random rate (%) | P value  |
|--------|--------|------------------|-----------------|----------|
| Hub    | 11, 10 | 70.0             | 43.1            | < 0.05   |
| SIM    | 13, 10 | 70.0             | 43.2            | < 0.04   |
| FFL    | 29, 21 | 61.9             | 42.9            | < 0.04   |
| Bi-fan | 38, 28 | 60.1             | 43.1            | < 0.03   |

In the second column, a pair of the numbers is given; the first number represents the total transcription factors (TFs) in each component (hub, SIM, FFL and bi-fan); the second number represents the total TFs which are transcript half-life (THL) mapped in each component. In the third column, the natural rate represents the observed fraction of the TFs having short transcript half-lives (sTHLs) in the THL-mapped TFs. When a TF's THL is equal to or shorter than 5.2 min, we say this TF is a sTHL TF. In the fourth column, the random rate represents the fraction of the sTHL TFs in the randomly sampled TFs. P value represents the probability.

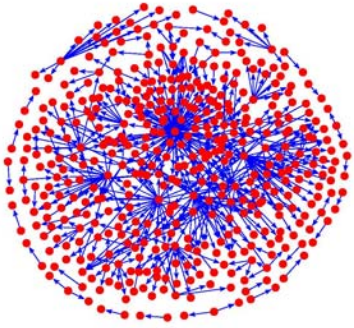

**a**

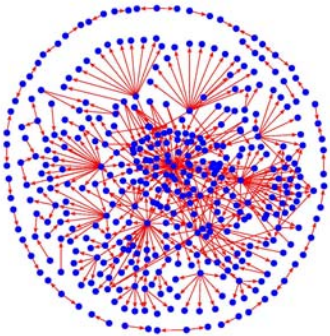

**b**

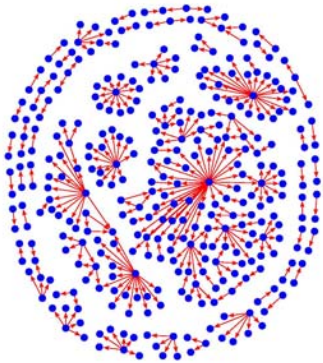

**c**

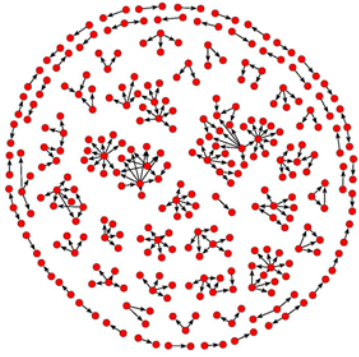

**d**

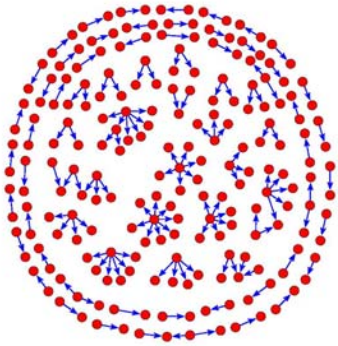

**e**

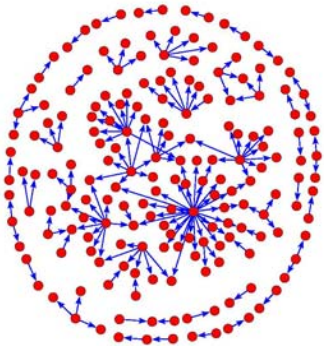

**f**

**Supplementary Figure 1: Targeted removal of network motif links. (a)** represents the static network. **(b)**, **(c)** and **(d)** represent the resulting networks after targeted removal of the links of FFLs, bi-fans and SIMs from the static network, respectively. **(e)** represents the resulting network after targeted removal of all the links of all three motif types from the static network. **(f)** represents the resulting network after randomly removing the same number of links as in (e) from the static network.

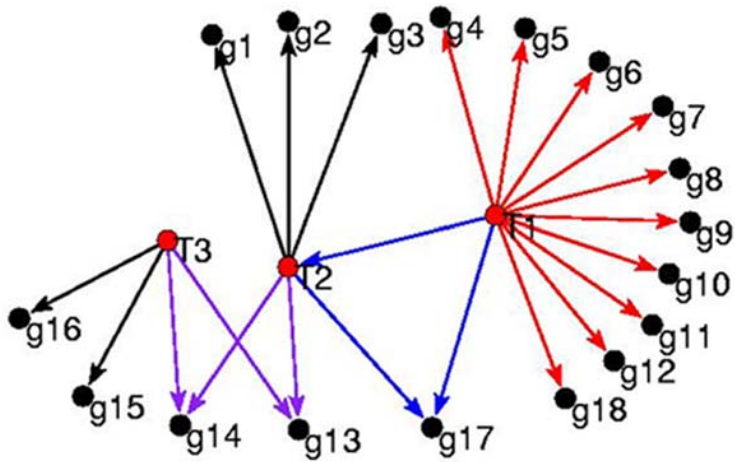

**Supplementary Figure 2: Network motifs and their links in the network.** This graph shows a portion of the network and the major motifs. Nodes in red and black represent transcription factors and target genes/operons, respectively. The links in blue and their connected nodes represent a feedforward loop (FFL) circuit. The links in purple and their connected nodes represent a bi-fan circuit. A Single Input Module (SIM) circuit consists of the links in red, the links of T1 -> T2 and T1->g17 and their connected nodes. A SIM circuit contains a TF and at least 9 target operons. It is noted that some circuits share a certain number of links. In this example, the SIM and the FFL circuits share links.

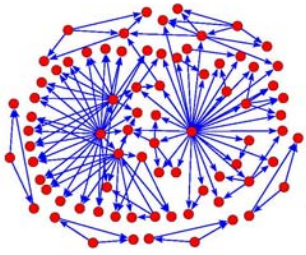

**a**

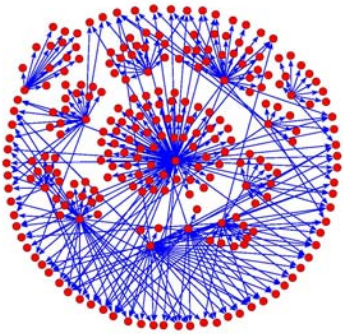

**b**

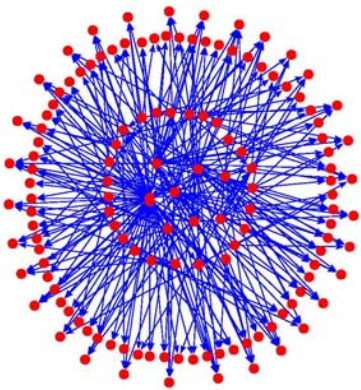

**c**

**Supplementary Figure 3: Self-organizing network motifs by sharing of transcription factors and operons. (a)**, **(b)** and **(c)** represent the networks generated by self-organizing feedforward loops, Single Input Modules and bi-fans, respectively. Red circles represent transcription factors (TFs) or operons; links with arrows represent the regulation relations between TFs and operons; the circles pointed at by arrows indicate the TFs or operons are regulated by the TFs, which send the arrow links. Motifs sharing same transcription factors or operons are self-assembled.